# The model of stress distribution in polymer electrolyte membrane

Vadim V. Atrazhev[c,d,z], Tatiana Yu. Astakhova[c], Dmitry V. Dmitriev[c,d], Nikolay S. Erikhman[c,d], Vadim I. Sultanov[d], Timothy Patterson[b,*] and Sergei F. Burlatsky[a,*]

[a] United Technologies Research Center, 411 Silver Lane, East Hartford, CT 06108, USA

[b] UTC Power, 195 Governor's Highway, South Windsor, CT 06074, USA

[c] Russian Academy of Science, Institute of Biochemical Physics, Kosygin str. 4, Moscow, 119334, Russia

[d] Science for Technology LLC, Leninskiy pr-t 95, 119313, Moscow, Russia

* Electrochemical Society Active Member.

[z] E-mail: vvatrazhev@deom.chph.ras.ru

**Abstract**

An analytical model of mechanical stress in a polymer electrolyte membrane (PEM) of a hydrogen/air fuel cell with porous Water Transfer Plates (WTP) is developed in this work. The model considers a mechanical stress in the membrane is a result of the cell load cycling under constant oxygen utilization. The load cycling causes the cycling of the inlet gas flow rate, which results in the membrane hydration/dehydration close to the gas inlet. Hydration/dehydration of the membrane leads to membrane swelling/shrinking, which causes mechanical stress in the constrained membrane. Mechanical stress results in through-plane crack formation. Thereby, the mechanical stress in the membrane causes mechanical failure of the membrane, limiting fuel cell lifetime. The model predicts the stress in the membrane as a function of the cell geometry, membrane material properties and operation conditions. The model was applied for stress calculation in GORE-SELECT®.

# 1. Introduction

A model of mechanical stress in the polymer electrolyte membrane (PEM) of a hydrogen/air fuel cell caused by transient operation of a fuel cell is presented in this paper. The fuel cells used in transportation applications can undergo as much as 150 load cycles per hour. For automotive and bus applications, the lifetime requirement is approximately 5,000 and 35,000 hours, respectively. This requires 750,000 cumulative load cycles for an automobile application and as many as 5 million cycles for a bus application. The large number of transient load cycles required for transportation applications will be challenging, with respect to membrane durability. The model, which predicts the membrane lifetime as a function of membrane mechanical properties, fuel cell geometry and operation conditions, can be used for optimization of fuel cell geometry and operation conditions for extension of membrane life-time.

The polymer membrane in the fuel cell is subjected to both chemical[1–5] and mechanical degradation[6]. The chemical degradation of the membrane is a result of chemical attack of the membrane by free radicals[4,7]. The free radicals[8] are generated in the membrane by oxygen reduction reaction catalyzed by Pt particles, which precipitate in the membrane during the fuel cell operation[9]. Oxygen and hydrogen are transported into the membrane through diffusion from the cathode and anode sides, respectively. Elevated temperature and low relative humidity (RH) accelerate the chemical degradation. The chemical degradation leads to decay of the membrane mechanical strength and accelerates the mechanical failure of the membrane. Mechanical failure of the membrane typically manifests itself through cracks in the membrane. Delamination between the polymer membrane and the electrodes can occur as well[10–16]. The origin of the mechanical failure is a mechanical stress induced in the membrane during the fuel cell operation.

It is generally accepted in the literature that the mechanical stress in the membrane caused by the repeated cycling of the membrane water content is the major factor of the membrane mechanical failure[17–19]. The load cycling causes the cycling of the inlet gas flow rate under constant utilization conditions. That leads to the cycling of air relative humidity (RH) in gas channels. At high RH, the membrane absorbs water and swells. At low RH, the membrane desorbs water and shrinks. The membrane is constrained in the fuel cell by by-polar plate through Gas Diffusion Layers (GDLs), which impedes dimensional change of the membrane. The swelling/shrinking of the constrained membrane causes mechanical stress in the membrane, which results in through-plane crack generation. It is experimentally observed that the variation of gas RH from 100% to 25% results in the membrane stress approximately equal to 2 MPa[20].

In the current work, we modeled the membrane in specific fuel cell design with Water Transport Plate (WTP). The WTPs are made of porous graphite with gas and coolant channels slotted in it. The functions of WTP are to humidify the inlet air and remove water generated in the cathode under fuel cell operation. The pores of WTP are filled by water under fuel cell operation and therefore the gas in the channel in WTP is humidified. However, the humidification of the inlet dry air is not instant and some fraction of the gas channel is under dry conditions. The length of dry region is controlled by the inlet gas flow rate and varies with variation of the flow rate, which results in membrane humidity cycling in this region. The length of the dry region for typical fuel cell design is approximately equal to 1 centimeter. The cracks and pinholes in degraded membrane are observed in UTRC experiments in the narrow region ( ~ 1 cm) near air inlet. However, crossover of the gases through these cracks causes performance loss of the whole cell.

Currently, the membrane lifetime in vehicle operational conditions is over ten thousand hours. This time is too long for full-time experiments on the membrane degradation. Therefore, accelerated tests under aggressive conditions, i.e. elevated temperature and low RH, are carried out[13,18,20,21] for study of the membrane degradation rate. The impact of the temperature and RH on the mechanical properties and lifetime of Nafion membranes was studied previously[22–24]. It was found that decreasing membrane water content and increasing the operation temperature decrease membrane lifetime.

Reinforced composite membranes were used to improve membrane mechanical properties and increase membrane lifetime[25,26]. The reinforced membrane consists of durable polymer reinforcement filled with proton conducting ionomer, e.g. Nafion®. The durable polymer reinforcement improves the mechanical strength of the membrane. For example, Penner et al. describe a Nafion membrane reinforced by porous polytetrafluoroethylene (PTFE)[26]. Reinforced membranes with PTFE incorporated into Nafion were also developed by DuPont[25]. New micro-reinforced polymer electrolyte membranes, GORE-SELECT® membranes, were developed by W.L. Gore & Associates[27–29].

The mechanical properties of both non-reinforced and reinforced membranes were studied by Tang and co-authors[30,31]. The experimental dependencies of Young's modulus, proportional limit and break stresses and dimensional change on temperature and humidity level are presented for non-reinforced Nafion®112 membrane[30]. The results indicate that the Young's modulus and the proportional limit stress of the non-reinforced Nafion®112 membrane decrease as humidity and temperature increase. Higher temperature leads to the lower break stress and the higher break strain. However, the humidity has little effect on the break stress and break strain. A critical set of material properties for the reinforced GORE-SELECT membrane is determined for a range of temperature and humidity levels by Tang et al.[31]. The swelling coefficient is also measured as function of temperature and humidity level. It was observed that the swelling coefficient of the reinforced GORE-SELECT membrane is approximately 4 times smaller than that for a non-reinforced membrane.

Constitutive models of the polymer membrane mechanical behavior in the fuel cell conditions are available in the literature. These models utilize finite element analysis of complex linear and non-linear problems using commercially available software. The models are developing towards complication of the cell geometry and the membrane material properties. The pioneer linear elastic model of membrane[19] was further developed to incorporate elasto-plasticity[17,19,31–36]. Both non-reinforced[17,32] and reinforced[31] membranes were modeled with elasto-plastic theory. The visco-plasticity model, which consists of elasto-plastic network in parallel with an elastic-viscous network, was realized in ref[33]. Recently nonlinear viscoelastic-viscoplastic model was developed in ref[37] but the model was not utilized for simulation of membrane stress in the fuel cell. Optimization of fuel cell design for performance and durability would require programming and debugging the equations of the model developed in ref[17,19,31–37]. In some cases complicated non-linear interaction of model parameters results in prediction of very high internal stresses that is close to the membrane yield stress. We believe that membrane creep should relax such high stresses. Moreover, the available models are applicable only for a typical cell design with solid (non-porous) bi-polar plates. However, less common design with porous water filled bi-polar plate (Water Transport Plate - WTP) provides important benefit for membrane durability and performance[38].

The objective of this work is to develop an analytical model of the stress distribution in the membrane in fuel cell design with porous WTP. The model predicts the stress in the membrane as a function of the cell geometry, material properties and operation conditions. The model requires the water distribution in the membrane as an input. In the cell with the WTP, the water distribution in the membrane is governed by RH distribution in the gas channels, which depends on the coordinate along and across the channels (see Figure 1). In this work, the water distribution in the membrane is calculated in two steps. As the first step, the 3D RH distribution in the cathode gas channel is modeled. Then the water distribution in the membrane is calculated by the solution of the diffusion equation with RH distribution at the cathode gas channel/GDL interface as a boundary condition.

The reinforced membrane is modeled as a three-layer composite, where a layer of reinforcement is located between two layers of ionomer. The geometry of the membrane is assumed to be plane, i.e. no buckling and wrinkling are taken into account. The 3D analytical solution of linear elasticity equations in the membrane is obtained in the thin membrane approximation. The membrane creep is taken into account in steady state approximation. The model is applied for calculation of the RH cycling induced stress in the GORE-SELECT membrane under the typical fuel cell load cycling conditions.

## 2. RH distribution in cathode gas channel

In this section, we calculate 3D RH distribution in the cathode gas channel of the fuel cell with Water Transport Plate (WTP). The RH distribution in the cathode gas channel is required for calculation of the water content in the membrane. The qualitative picture of the gas flows in the fuel cell is shown in Figure 1. Cross-flow flow-field is typically used, when the anode and cathode channels are oriented in perpendicular directions. A Membrane Electrode Assembly (MEA) in the fuel cell is located between two GDLs and fixed between two WTPs. The WTPs are made of porous graphite and the pores are filled by water under fuel cell operation. The gas channels are machined in the WTP. The wet hydrogen is supplied through the anode channel. Therefore, the hydrogen RH is 100% in all parts of the anode channel. The dry ambient air is supplied through the cathode channel. The function of the WTP is to humidify the inlet ambient air through evaporation of the liquid water from the WTP pores and diffusion of water vapor into the gas stream. However, the humidification is not instant. The dry inlet gas gradually becomes saturated as it flows along the gas channel as depicted in Figure 1. The gas humidity level gradually increases along the cathode channel and reaches 99% RH at the distance ~2 cm from the gas inlet. At fixed coordinate along the channel, $y$, the RH has its minimum at the middle point of the channel and reaches 100% RH in the area under the WTP ribs. Alternation of the gas humidity under the channels and ribs drives the alternation of water content $\lambda(x, y, z)$ in the membrane.

We divided calculation of $\lambda(x, y, z, t)$ into two steps taking advantage of scale separation. The thickness of the GDL that separates the MEA from the gas channels is of the order of 100 μm (Figure 2). That is an order of magnitude smaller than the gas channel thickness (approximately 1 mm). At the first step, we calculate 3D vapor distribution, $C(x, y, z)$, in the cathode gas channel near air inlet. This distribution is calculated numerically and the numerical solution is approximated by analytical expression to reduce computational time and make results more transparent. At the second step, 3D $\lambda(x, y, z, t)$ distribution in the membrane was calculated using vapor concentration at the gas channel/GDL interface, $C(H_{ch}, y, z)$, as a boundary condition for diffusion equation for the water in the membrane and neglecting $\lambda$ derivatives with respect to $y$ and $z$. The resulting 3D water distribution in the membrane parametrically depends on $y$ and $z$ coordinates through the boundary conditions.

The typical flow rate cycling protocol is modeled as a periodic stepwise variation of load, $j$, that consists of two time steps with low current density, $j = j_{min}$, and high current density, $j = j_{max}$. The typical cycle period ($T_{cyc} \sim 1$ min) is much longer than the time required to reach the steady state flow and RH distribution in the gas channel. Therefore, we calculate the steady state vapor distribution in the gas channel at fixed gas flow rate and model the flow cycling as a sequential switching of the steady state flows with the different flow rates.

The steady state mass balance equation for the water vapor concentration, $C$, in the gas channel is

$$\mathrm{div}(-D\vec{\nabla}C + \vec{V}C) = 0 \tag{1}$$

Here $\vec{V}$ is the gas velocity in the channel and $D$ is the vapor diffusion coefficient. Only $y$ component of the gas velocity, $V_y$, is non-zero in the channel. Taking advantage of the fact that the convective vapor flux in $y$ direction, $V_y C$, is much larger than the diffusion vapor flux in the same direction, $-D\partial C/\partial y$, we neglect the term $-D\partial C/\partial y$ in Equation (1). That results in the following equation for the vapor concentration

$$-D\left(\frac{\partial^2 C}{\partial x^2} + \frac{\partial^2 C}{\partial z^2}\right) + V_y(x,z)\frac{\partial C}{\partial y} = 0 \tag{2}$$

We introduce new dimensionless variables: $x' = x/H_{ch}$, $z' = z/L_{ch}$, $y' = y/\xi$. The scaling parameter $\xi = J_0/D$ is proportional to the air humidification length in the gas channel. Here $J_0 = V_0 L_{ch} H_{ch}$ is the total gas flow in the channel and $V_0 = \frac{1}{L_{ch} H_{ch}}\int V_y(x,z)dxdz$ is the mean gas velocity in the channel. Substituting new variables into Equation (2), we obtain

$$-\left(\frac{L_{ch}}{H_{ch}}\frac{\partial^2 C}{\partial x'^2} + \frac{H_{ch}}{L_{ch}}\frac{\partial^2 C}{\partial z'^2}\right) + v_y(x,z)\frac{\partial C}{\partial y'} = 0 \tag{3}$$

Here $v_y(x,z) = V_y(x,z)/V_0$ is normalized gas velocity.

The boundary conditions for Equation (3) are derived below. We assume that the concentration of the vapor at the channel walls is equal to the concentration of saturated vapor at the temperature of the wall, $C_{sat}$. We assume here that high heat conductivity of the carbon (WTP material) leads to the uniform temperature distribution in the WTP. The vapor flux into the gas channel consists of the flux from WTP water filled pores (channel walls) and the flux from the cathode. The vapor flux from the cathode is equal to water generation rate due to electrochemical reaction in the cathode. In fuel cell with porous WTP the partial water vapor pressure saturates in relatively short humidification zone located at the channel inlet due to fast water evaporation form the channel walls. We assume that in humidification zone, the vapor flux from the cathode into the channel through the GDL is much lower than the vapor flux from the channel walls. Comparison of obtained numerical solution for the vapor flux from the channel walls with water generation rate in the cathode at current density up to 2 A/cm$^2$ justifies our assumption. Therefore, we adopt the zero flux boundary condition at the GDL/channel interface. The above assumptions result in the following boundary conditions for Equation (3):

$$C(0, y, z) = C_{sat}; \quad \frac{\partial C(H_{ch}, y, z)}{\partial x} = 0;$$
$$C(x, y, 0) = C_{sat}; \quad C(x, y, L_{ch}) = C_{sat}; \quad C(x, 0, z) = C_{in} \qquad (4)$$

Here $C_{in}$ is the vapor concentration in the inlet gas. Equation (3) for the vapor concentration in the channel requires the gas flow distribution in the channel, $V_y(x, z)$, as an input. The flow distribution in the channel is governed by the steady state Navier-Stokes equation:

$$\rho(\vec{V}\nabla)\vec{V} = -\nabla P_{gas} + \mu \Delta \vec{V} \qquad (5)$$

Only y-component of the gas velocity, $V_y$, is not equal to zero in Equation (5). We assume that the gas is incompressible and that the gas velocity is independent on y-coordinate, i.e. $\Delta V_y = \partial^2 V_y / \partial x^2 + \partial^2 V_y / \partial z^2$. The total gas pressure, $P_{gas}$, varies only in y-direction along the channel, i.e. $\nabla P_{gas} = \partial P_{gas}/\partial y$. The Reynolds number in the gas channel is small (Re ~ 10) for the typical gas channel geometry and the gas flow rate. Therefore, the flow in the channel is laminar and we neglect the term $\rho(\vec{V}\nabla)\vec{V}$ in Equation (5). Taking advantage of the above assumptions and substituting dimensionless variables $x'$, $z'$ and $v_y$ into Equation (5) we obtain:

$$\left(\frac{L_{ch}}{H_{ch}}\frac{\partial^2 v_y}{\partial x'^2} + \frac{H_{ch}}{L_{ch}}\frac{\partial^2 v_y}{\partial z'^2}\right) = \frac{L_{ch}H_{ch}}{\mu V_0}\frac{\partial P_{gas}}{\partial y} \tag{6}$$

Assuming $\partial P_{gas}/\partial y = const$ and normalizing both sides of the Equation (7) by $\frac{L_{ch}H_{ch}}{\mu V_0}\frac{\partial P_{gas}}{\partial y}$ we obtain the following 2D equation for renormalized velocity distribution $\tilde{v}_y(x,z)$

$$\left(\frac{L_{ch}}{H_{ch}}\frac{\partial^2 \tilde{v}_y}{\partial x'^2} + \frac{H_{ch}}{L_{ch}}\frac{\partial^2 \tilde{v}_y}{\partial z'^2}\right) = 1 \tag{7}$$

We use flow-slip boundary condition at the walls for Equation (7)

$$\tilde{v}_y(0,z) = 0; \quad \tilde{v}_y(H_{ch},z) = 0;$$
$$\tilde{v}_y(x,0) = 0; \quad \tilde{v}_y(x,L_{ch}) = 0; \tag{8}$$

The normalized gas velocity, $v_y(x,z)$, is calculated from solution of Equation (7) by the equation

$$v_y(x,z) = \frac{L_{ch}H_{ch}\tilde{v}_y(x,z)}{\int \tilde{v}_y(x,z)dxdz} \tag{8a}$$

Substituting $v_y(x,z)$ into Equation (3) and solving this equation numerically with the boundary conditions (4), we obtain the vapor distribution in the gas channel, $C(x, y, z)$.

The vapor distribution at the GDL/gas channel interface is required for calculation of water content in the membrane. To accelerate the model we approximated the numerical solution for the vapor concentration by an analytical expression. The additional advantage of using the analytical expression is that it is more transparent and can be easily used for the stack design optimization. We approximate the vapor concentration at the GDL/gas channel interface by the following expression:

$$C(x = H_{ch}, y, z) = C_{in} + (C_{sat}(T_{cool}) - C_{in})f(y,z)$$
$$f(y,z) = 1 - \exp\left(-\alpha\frac{y}{\xi}\right)\left(1 - 4\left(\frac{z}{L_{ch}} - \frac{1}{2}\right)^2\right) \tag{9}$$

Here $\alpha$ is a fitting parameter, which is obtained by fitting the numerical solution by Equation (9). The numerical solution for the vapor concentration at the GDL/gas channel interface for $C_{in} = 0$ and $C_{sat} = 1$ and the function $f(y, z)$ are shown in Figure 3.

From Equation (7) it follows that $v_y(x,z)$ depends only on the ratio $L_{ch}/H_{ch}$ for the rectangular channel cross-section. Equation (3) also includes only one explicit geometrical parameter $L_{ch}/H_{ch}$. Therefore, the vapor distribution in the gas channel in dimensionless coordinates depends only on one geometrical parameter, $L_{ch}/H_{ch}$, and the numerical parameter $\alpha$ depends only on the $L_{ch}/H_{ch}$ ratio as well. We calculated numerically the vapor distribution in the gas channel for several values of $L_{ch}/H_{ch}$ and fitted these numerical solutions by Equations (9), using $\alpha$ as a fitting parameter. The values of the parameter $\alpha$ for the several values of $L_{ch}/H_{ch}$ are plotted in the Figure 4. The parameter $\alpha$ slowly depends on $L_{ch}/H_{ch}$ for typical values of $L_{ch}/H_{ch}$, as indicated in the Figure 4. Therefore, we use the average value of $\alpha = 8$ in the model.

As a result, we obtain the analytical expression (9) for the distribution of the vapor concentration at the cathode gas channel/GDL interface near the gas inlet. This distribution depends only on one lumped parameter $\xi = V_0 L_{ch} H_{ch}/D$. The humidification length of the vapor in the channel is $L_{hum} = \xi/\alpha$. Thereby, the humidification length is proportional to the gas flow rate in the channel. The gas flow rate is proportional to the cell load under the constant utilization conditions. Therefore, the humidification length is proportional to the cell load. If the cell load varies by the order of magnitude under operation conditions, the humidification length also varies by the order of magnitude. That results in substantial variation of the membrane water content close to the air inlet.

## 3. Water content in membrane

In this section, we calculate the water content in the membrane using the vapor distribution in the air gas channel. The water content in the membrane, $\lambda(x, y, z, t)$, is governed by the vapor distribution at the cathode channel/GDL interface, $C_C(y, z, t)$, at the anode channel/GDL interface, $C_A$, and the concentration drop in the GDLs, as depicted in Figure 5. The steady-state vapor distribution at the GDL/gas channel interface in the cathode side, calculated in the previous section, is used as a boundary condition. The vapor concentration in the anode channel, $C_A$, is assumed to be constant and equal to the concentration of the saturated vapor. The water content in the membrane is calculated in two steps. At first step, we calculate steady state concentrations at the membrane/GDL interfaces, $C_1(y, z)$ and $C_2(y, z)$. In this step, we replace the actual local diffusion coefficient in the membrane that depends on the local $\lambda$ with its value averaged over the humidity cycling. The vapor concentrations at the membrane/GDL interfaces were calculated from the linear diffusion equation resulting from the above approximation. The flux conservation conditions at the membrane/GDL interfaces were used. At the second step, we calculate $\lambda(x, y, z, t)$ through solving the non-linear 1D diffusion equation with the local diffusion coefficient in the membrane that depends on the local $\lambda$ with the water vapor concentrations at the membrane/electrode interfaces, $C_1(y, z, t)$ and $C_2(y, z, t)$ (see Figure 5), as the boundary conditions. The calculated water content in the membrane depends on $y$ and $z$ coordinates and time $t$ through $C_1(y, z, t)$ and $C_2(y, z, t)$.

The steady state approximation for the vapor transport in the GDLs and the membrane can be used because diffusion times in the GDL and the membrane are much smaller than the cycling period, $T_{cyc}$. Diffusion time in the membrane is $t_M = L_M^2/D_M \sim 1\,\text{s}$ and in the GDL it is $t_{GDL} = L_{GDL}^2/D_{GDL} \sim 10^{-3}\,\text{s}$. The GDL thickness is much larger than the thickness of the electrode and, therefore, the diffusion resistance of the GDL is much larger than that of the electrode and we neglect the diffusion resistance of the electrodes. We also model the water generation/consumption in the electrodes as the water source/sink at the membrane/GDL interface. The balance equations for the vapor fluxes at the membrane/anode GDL and the membrane/cathode GDL are:

$$\begin{aligned} J_M - J_{AGDL} &= -\frac{j}{F} \\ J_{CGDL} - J_M &= \frac{3j}{2F} \end{aligned} \qquad (10)$$

Here $J_M$, $J_{AGDL}$ and $J_{CGDL}$ are the water diffusion fluxes through the membrane, anode and cathode GDLs, correspondingly. The right hand side of the first Equation (10) is the water sink in the anode, which is caused by a drag of approximately one water molecule by one proton through the membrane. The right hand side of the second Equation (10) is the water source, which is caused by the water generation in the cathode and by the release of one water molecule dragged by the proton through the membrane.

Substituting the equations for the diffusion fluxes into Equation (10) and utilizing the average water diffusion coefficient in the membrane, $D_M$, we obtain:

$$\frac{D_{GDL}}{L_{GDL}}(C_1 - C_A) - B\frac{D_M}{L_M}(C_2 - C_1) = -\frac{j}{F}$$
$$B\frac{D_M}{L_M}(C_2 - C_1) - \frac{D_{GDL}}{L_{GDL}}(C_C - C_2) = \frac{3j}{2F}$$
(11)

Here $B = \frac{\rho}{EW}\frac{d\lambda_{eq}(C)}{dC}$, where $\lambda_{eq}(C)$ is obtained from the experimental equilibrium water uptake isotherm [39], $\rho = 1.5$ g cm$^{-1}$ and EW = 1100 g mol$^{-1}$ is the Nafion equivalent weight. We assume a thermodynamic equilibrium of the water in the membrane with the water vapor at the membrane/GDL interface. The solution of Equations (11) is

$$C_1 = \frac{(a+b)C_A + bC_C}{a + 2b} - \frac{j}{F}\frac{(3/2)(a+b) - b}{a(a + 2b)}$$
$$C_2 = \frac{(a+b)C_C + bC_A}{a + 2b} - \frac{j}{F}\frac{(3/2)b - (a+b)}{a(a + 2b)}$$
(12)

Here $a = \frac{D_{GDL}}{L_{GDL}}$, $b = B\frac{D_M}{L_M}$.

The RH = 100% in the anode gas channel and $C_A = C_{sat}(T_{cool})$. The vapor concentration in the cathode gas channel, $C_C$, is calculated by Equation (9). The humidification length $\xi = V_0 L_{ch} H_{ch}/D$ depends on the gas velocity in the channel, which in turn depends on the current density, $j$. The gas velocity under the constant utilization conditions is calculated from the mass balance equation:

$$V_0(j) = \frac{jS_{pl}}{4FC_{O_2} n_{ch} H_{ch} L_{ch} U_{O_2}}$$
(13)

Here $S_{pl}$ is the platform active surface area, $C_{O_2}$ is the inlet oxygen concentration, $U_{O_2}$ is the oxygen utilization, $n_{ch}$ is a number of the channels in the platform. The vapor concentrations at the membrane/GDL interfaces, $C_{M/AGDL}(y, z, t)$ and $C_{M/CGDL}(y, z, t)$ are obtained from the steady state concentrations for the low and high load cycle steps ($j = j_{min}$ and $j = j_{max}$).

$$\begin{aligned} C_{M/CGDL}(y,z,t) &= C_2(y,z,j=j_{min}), \quad C_{M/AGDL}(y,z,t) = C_1(y,z,j=j_{min}), \quad 0 < t < T_{cyc}/2 \\ C_{M/CGDL}(y,z,t) &= C_2(y,z,j=j_{max}), \quad C_{M/AGDL}(y,z,t) = C_1(y,z,j=j_{max}), \quad T_{cyc}/2 < t < T_{cyc} \end{aligned} \quad (14)$$

The vapor concentrations $C_1$ and $C_2$ are used as the boundary conditions for the non-linear transient water diffusion equation in the membrane:

$$\frac{\partial \lambda}{\partial t} + \frac{\partial}{\partial x}\left(D(\lambda)\frac{\partial \lambda}{\partial x}\right) = 0 \quad (15)$$

The dependence the water diffusion coefficient on $\lambda$ is taken from [40].

Thermodynamic equilibrium at the membrane/gas interface results in the following boundary conditions for Equation (15)

$$\begin{aligned} \lambda(0) &= \lambda_{eq}\left(C_{M/CGDL}(y,z,t)\right) \\ \lambda(L_M) &= \lambda_{eq}\left(C_{M/AGDL}(y,z,t)\right) \end{aligned} \quad (16)$$

Equation (15) with the boundary conditions (16) is solved numerically. Water content in the membrane depends on the in-plane coordinates $y$ and $z$. The maximal variation of the water content is located near the gas inlet (assume $y = 5$ mm) in the middle of the channel ($z = L_{ch}/2$). The calculated water content as a function of the through-plane coordinate in this point for the wet and dry conditions are shown in Figure 6. The wet conditions correspond to the current density $j_{min} = 100$ mA cm$^{-2}$, and the dry conditions correspond to the current density $j_{max} = 1000$ mA cm$^{-2}$.

The variation of the water content at the cathode side of the membrane $\Delta\lambda \sim 4$ and at the anode side of the membrane $\Delta\lambda \sim 2$ was found when the load is cycled between $j_{min} = 100$ mA cm$^{-2}$ and $j_{max} = 1000$ mA cm$^{-2}$. The variation of the water content at the anode side of the membrane is a result of the GDL diffusion limitation, which causes decrease of vapor concentration at the anode GDL/membrane interface relative to the vapor concentration in the anode gas channel.

## 4. Stress distribution in membrane

In this section, we calculate the mechanical stress in the membrane induced by the variation of the water content in the membrane. The ionomer membranes currently used for PEM fuel cells contain polymeric reinforcement, which improves mechanical strength of the membrane. We model the reinforced membrane as a three-layer composite with the known mechanical properties of the each layer and calculate the stress distribution in the reinforced membrane.

The membrane is mechanically constrained in fuel cell between the WTPs by the GDLs. We focus on a stack with cross-flow flow-field where the anode and the cathode gas channels are directed in perpendicular directions, as shown in Figure 7. We assume that the membrane is constrained by the anode and the cathode WTP flow-field ribs, and there is overlap of both ribs, as indicated by a set of squares in Figure 7a. This assumption is supported by experimental observation of a checkerboard pattern on the membrane. We assume that in these regions the membrane cannot slip along the GDL surface. The GDL consist of highly porous material with porosity approximately equal to 0.8. That results in high compressibility of the GDL, and we assume that the membrane can easyly change size in $x$ direction during swelling without generation of large mechanical stress in the GDL. Therefore, we assume that normal component of the mechanical stress at the membrane/GDL interface, $\sigma_{xx}$, is zero.

We use the simplified model geometry to obtain the analytical solution for the stress in the membrane. The membrane is permanently wet in the regions under the cathode WTP ribs because the membrane is close to the wet WTP in these regions. The dry regions are located under the cathode channels and extended along the channels as shown in Figure 1. The typical humidification length in the cathode gas channel approximately equals 1 cm, which is by the order of magnitude larger than the channel thickness, $L_{ch} \sim 1$ mm. Therefore, all derivatives of the deformation and stress with respect to coordinate $y$ are by the order of magnitude smaller than the derivatives with respect to coordinate $z$ and we neglect them in the following equations. The displacement of the membrane in $z$ direction is a periodic function. Taking advantage from the assumed displacement symmetry, we conclude that the membrane displacement in $z$ direction equals zero under the middle of the cathode ribs. That enables the further simplification of the model as illustrated in Figs. 7a and 7b.

In-plane stress components in the membrane, $\sigma_{yy}$ and $\sigma_{zz}$, cause through-plane crack formation and the membrane failure. Therefore, we focus on calculation of the in-plane component of the stress tensor. Taking advantage of the fact that membrane thickness (~ 20 μm) is much smaller than the membrane size in $y$ and $z$ directions, we calculate the in-plane components of the stress without calculation of non-diagonal components of the stress tensor, which have no influence on membrane lifetime.

As the first step, we calculate the membrane elastic response to an instant change of the water content in the membrane, $\Delta\lambda(x,y,z) = \lambda(x,y,z) - \lambda_0$. The diagonal components of the elastic deformation and the stress of the membrane are governed by equations of the linear elasticity:

$$\sigma_{xx} = \frac{E\nu}{(1+\nu)(1-2\nu)}\left(\varepsilon_{xx} + \varepsilon_{yy} + \varepsilon_{zz}\right) + \frac{E}{(1+\nu)}\varepsilon_{xx} - \frac{E\alpha}{3(1-2\nu)}\Delta\lambda(x,z)$$
$$\sigma_{yy} = \frac{E\nu}{(1+\nu)(1-2\nu)}\left(\varepsilon_{xx} + \varepsilon_{yy} + \varepsilon_{zz}\right) + \frac{E}{(1+\nu)}\varepsilon_{yy} - \frac{E\alpha}{3(1-2\nu)}\Delta\lambda(x,z) \quad (17)$$
$$\sigma_{zz} = \frac{E\nu}{(1+\nu)(1-2\nu)}\left(\varepsilon_{xx} + \varepsilon_{yy} + \varepsilon_{zz}\right) + \frac{E}{(1+\nu)}\varepsilon_{zz} - \frac{E\alpha}{3(1-2\nu)}\Delta\lambda(x,z)$$

Here $\varepsilon_{ij}$ is the deformation tensor, $\sigma_{ij}$ is the stress tensor, $E$ is the Young modulus and $\nu$ is the Poisson coefficient, $\alpha$ is the swelling coefficient of the membrane. At the equilibrium water content in the membrane, $\lambda_0$, the membrane is not stressed. The Young modulus and the swelling coefficient depend on the through-plane coordinate $x$ because $E$ and $\alpha$ of the reinforcement differs from that of the Nafion. The Young modulus also depends on the water content that results in implicit dependence $E(x)$. Taking into account that the deformation in $y$ direction $\varepsilon_{yy} = 0$ and the stress in $x$ direction $\sigma_{xx} = 0$, we obtain from Equations (17)

$$\sigma_{yy} = \frac{E}{1-\nu}\left[\frac{\nu}{1+\nu}\varepsilon_{zz} - \frac{\alpha\Delta\lambda}{3}\right]$$
$$\sigma_{zz} = \frac{E}{1-\nu}\left[\frac{1}{1+\nu}\varepsilon_{zz} - \frac{\alpha\Delta\lambda}{3}\right] \quad (18)$$

Averaging the second Equation (18) over *x* we obtain

$$\langle \sigma_{zz} \rangle_x = \frac{1}{1-\nu} \left[ \frac{1}{1+\nu} \langle E \rangle_x \varepsilon_{zz} - \frac{\langle E\alpha\Delta\lambda \rangle_x}{3} \right] \tag{19}$$

We assume that the GDLs prevent the membrane buckling and wrinkling. Absence of the buckling or wrinkling leads to independence of the in-plane deformation of the membrane, $\varepsilon_{zz}$, on through-plane coordinate *x*. In the opposite case, the small difference between $\varepsilon_{zz}$ at the cathode and at the anode sides would lead to the large deformation of the membrane in *x* direction and the deviation of the membrane shape from the flat one. Averaging Equation (19) over *z* we obtain

$$\langle \sigma_{zz} \rangle_{xz} = \langle \sigma_{zz} \rangle_x = \frac{1}{1-\nu^2} \langle E \rangle_{xz} \langle \varepsilon_{zz} \rangle_z - \frac{\langle E\alpha\Delta\lambda \rangle_{xz}}{3(1-\nu)} \tag{20}$$

Here we use approximate uncoupling of the average value $\langle \langle E \rangle_x \varepsilon_{zz} \rangle_z \approx \langle E \rangle_{xz} \langle \varepsilon_{zz} \rangle_z$. The membrane is constrained between the WTP ribs and the total size of the membrane in *z* direction is fixed, *i.e.* the average deformation $\langle \varepsilon_{zz} \rangle_z = 0$. Taking advantage of the fact that $\langle \varepsilon_{zz} \rangle_z = 0$ we obtain from Equation (20)

$$\langle \sigma_{zz} \rangle_x = -\frac{\langle E\alpha\Delta\lambda \rangle_{xz}}{3(1-\nu)} \tag{21}$$

Substituting Equation (21) into (19) we obtain the equation for $\varepsilon_{zz}$

$$\varepsilon_{zz} = \frac{1+\nu}{3\langle E \rangle_{xz}} \left[ \langle \alpha E\Delta\lambda \rangle_x - \langle \alpha E\Delta\lambda \rangle_{xz} \right] \tag{22}$$

Substituting Equation (22) into (18) we obtain the final equations for the elastic stress in the membrane caused by the instant change of the water content $\Delta\lambda(x,y,z)$:

$$\begin{aligned}
\sigma_{zz}(x,y,z) &= -\frac{E\alpha}{3(1-\nu)}\Delta\lambda + \frac{E}{3(1-\nu)} \frac{\left( \langle E\alpha\Delta\lambda \rangle_x - \langle E\alpha\Delta\lambda \rangle_{xz} \right)}{\langle E \rangle_{xz}}, \\
\sigma_{yy}(x,y,z) &= -\frac{E\alpha}{3(1-\nu)}\Delta\lambda + \frac{E\nu}{3(1-\nu)} \frac{\left( \langle E\alpha\Delta\lambda \rangle_x - \langle E\alpha\Delta\lambda \rangle_{xz} \right)}{\langle E \rangle_{xz}},
\end{aligned} \tag{23}$$

Substituting the water content in the membrane, calculated in previous section, into equations (23) we obtain elastic in-plane stress in the membrane. The Young moduli of the reinforced and non-reinforced membranes are taken from the literature[30,31]. The swelling coefficient of the Nafion was calculated using water balance in Nafion membrane and assuming that water and ionomer are incompressible. Swelling coefficient of reinforcement is assumed equal to zero. This assumption is confirmed by the experimentally measured dimensional change of the reinforced membrane [31] which is 6 times smaller than that of the Nafion 112 membrane at $T = 85°C$.

The elastic stress in the membrane was calculated from Equations (23) for the load cycling with the maximal load $j_{max} = 1000$ mA cm$^{-2}$ and the minimal load $j_{min} = 100$ mA cm$^{-2}$. The oxygen utilization was assumed 50% in this calculation. The calculated water content in the membrane near the air inlet at $z = L_{ch}/2$ for these conditions is shown in Figure 6. Substituting $\Delta\lambda(x,y,z) = \lambda(x,y,z) - \lambda_0$ at $y = 0.5$ cm into Equations (22) we calculate the elastic stress in the membrane. The through-plane distribution of the elastic stress at the middle of the air channel ($z = L_{ch}/2$) is shown in Figure 8. The stress in the reinforcement is negative because the reinforcement swelling coefficient, $\alpha_R$, is assumed equal to zero. The Nafion shrinks at the low water content, which results in a tensile (positive) stress in the Nafion. The Nafion acts on the reinforcement and causes a compressive (negative) stress in the reinforcement at the low water content. The tensile stress in the Nafion at the cathode side of the membrane reaches 8 MPa, which is close to the yield stress of the Nafion. Thereby, such a large stress leads to fast plastic deformation of the membrane. Moreover, the creep of the polymer under the low level of stress that is below the yield strength of the polymer causes the relaxation of the membrane stress. It was shown experimentally in [41] that dry Nafion subjected to 1.55 MPa tensile stress during 5 hours at 70°C elongates on approximately 50% due to creep. The Nafion at 65% RH elongates on approximately 10% under the same conditions. Since the membrane in the fuel cell is subjected to hydration-induced stress during the hundreds of hours, the irreversible deformation of the membrane is substantial and the membrane creep should be incorporated into the model.

The membrane creep under the long-term stress results in irreversible elongation or compression of the membrane, *i.e.* it results in a new equilibrium length of the membrane. The dimensions of the membrane in the fuel cell are fixed by WTP through GDLs and cannot be changed. The change of the water content in the membrane causes mechanical stress. However, the membrane creep relaxes the stress with time[19]. In our model equilibrium water content, $\lambda_{eq}$, corresponds to an unstressed membrane. Therefore, if the water content is fixed at some value $\lambda$ the equilibrium water content relaxes to $\lambda$ due to membrane creep.

The period of the stress variation in the membrane is equal to the period of the load cycling, which is of the order of 100 s. In this paper, we consider the cycling with the period, $T_{cyc}$, much smaller than the average relaxation time $\tau$ of the polymer, which is higher than $10^3$ s. Therefore, the large irreversible deformation cannot occur during one cycling period. The appreciable irreversible deformation of the membrane due to creep occurs at the time interval of the order of $\tau$, *i.e.* during the large number of cycles. Under cycling conditions, the equilibrium water content is a slow varying function. We assume that at steady-state approximately equal to the local water content averaged over time

$$\lambda_{eq}(x,y,z) = \langle \lambda \rangle_t = \frac{1}{T_{cyc}} \int_0^{T_{cyc}} \lambda(x,y,z,t) dt \qquad (24)$$

where $\lambda(x, y, z, t)$ is a solution of Equation (15) with a boundary conditions (16).

The deviation of the local water content, $\lambda$, from the equilibrium value results in elastic stress in the membrane. Substituting $\Delta\lambda = \lambda(x,y,z,t) - \langle \lambda \rangle_t$ into Equations (23), we obtain the equations for the steady-state stress in the membrane after a cycling time $\gg \tau$:

$$\begin{aligned}\sigma_{zz}(x,y,z) &= -\frac{E\alpha}{3(1-\nu)}(\lambda - \langle\lambda\rangle_t) + \frac{E}{3(1-\nu)} \frac{\left(\langle E\alpha(\lambda - \langle\lambda\rangle_t)\rangle_x - \langle E\alpha(\lambda - \langle\lambda\rangle_t)\rangle_{xz}\right)}{\langle E\rangle_x}, \\ \sigma_{yy}(x,y,z) &= -\frac{E\alpha}{3(1-\nu)}(\lambda - \langle\lambda\rangle_t) + \frac{E\nu}{3(1-\nu)} \frac{\left(\langle E\alpha(\lambda - \langle\lambda\rangle_t)\rangle_x - \langle E\alpha(\lambda - \langle\lambda\rangle_t)\rangle_{xz}\right)}{\langle E\rangle_x},\end{aligned} \qquad (25)$$

The viscoelastic stress in the membrane under the load cycling with $j_{max} = 1000$ mA cm$^{-2}$, $j_{min} = 100$ mA cm$^{-2}$ and the oxygen utilization 50% was calculated. The in-plane stress in the membrane as a function of the through-plane coordinate, $x$, at the middle of the air channel ($z = L_{ch}/2$) at $y = 0.5$ cm is shown in Figure 9.

The feature of the viscoelastic stress in the membrane is alternating in time of the tensile and compressive stress in the local coordinate, as shown in Figure 9. The sign of the water content deviation from $\lambda_{eq}$ determines the sign of the stress. The local equilibrium water content, $\lambda_{eq}(x)$, lies in the interval between the local $\lambda_{min}(x)$ and $\lambda_{max}(x)$. Thereby, the local $\lambda$ deviates in both sides, above and below, from the local $\lambda_{eq}$ during the load cycling. The deviation of the local $\lambda$ below the local $\lambda_{eq}$ causes the Nafion shrinking and a tensile (positive) stress in the Nafion. A compressive (negative) stress in the reinforcement appears at the same time. The deviation of the local $\lambda$ above the local $\lambda_{eq}$ causes the Nafion swelling and a compressive (negative) local stress. A tensile (positive) stress in the reinforcement appears at the same time. The stress relaxation under the plastic deformation reduces the maximal stress in the Nafion by approximately 4 times at the cathode/membrane interface. That substantially increases the estimate for the membrane lifetime because the experiments with the membrane failure under the stress cycling indicate that the membrane lifetime exponentially depends on the amplitude of the applied stress.

## 5. Conclusions

The analytical model of the stress distribution in the PEM membrane of the hydrogen/oxygen fuel cell under the load cycling conditions is developed. The mechanical stress in the membrane causes the through-plane cracks propagation and the membrane mechanical failure. The membrane lifetime as a function of applied stress can be measured in the out-of-cell experiment. To predict the membrane lifetime in the cell the knowledge of the in-cell stress of the membrane is required. However, the experimental measurement of the in-cell stress of the membrane is a challenging problem. Thereby, the model of the stress distribution in the membrane under the fuel cell operation conditions is required for prediction of the membrane in-cell lifetime.

The fuel cell with Water Transport Plate (WTP) and the reinforced membrane is under consideration. The developed model includes the plastic deformation of the membrane. The equations for the elastic-plastic stress in the reinforced membrane near the air inlet caused by the local variation of the gas relative humidity (RH) are derived. The in-plane stress distribution in the membrane as a function of the through-plane coordinate is calculated for the typical load cycling conditions (Figure 9). The local gas RH variation near the air inlet is a result of the load cycling. The model predicts that the stress at the membrane/cathode interface can reach 2 MPa. The comparison of the viscoelastic stress in the membrane with the elastic response (Figure 8) under the same conditions shows that the membrane creep reduces the maximal stress at the membrane/cathode interface approximately by 4 times. The membrane creep leads to alternating in time of the tensile and compressive stress in the local point of the membrane. The model predicts the lower stress in the reinforcement than that in the ionomer. In addition, the sign of the stress in the reinforcement is opposite to the sign of the stress in the ionomer, i.e. the tensile stress in the ionomer is accompanied by the compressive stress in the reinforcement. We speculate that such feature of the stress distribution in the reinforced membrane leads to deceleration of the crack propagation through the reinforced membrane and extension of the membrane lifetime.

**Acknowledgment**

This work was supported by the US Department of Energy under grant DE-EE0000468.

Figure 1

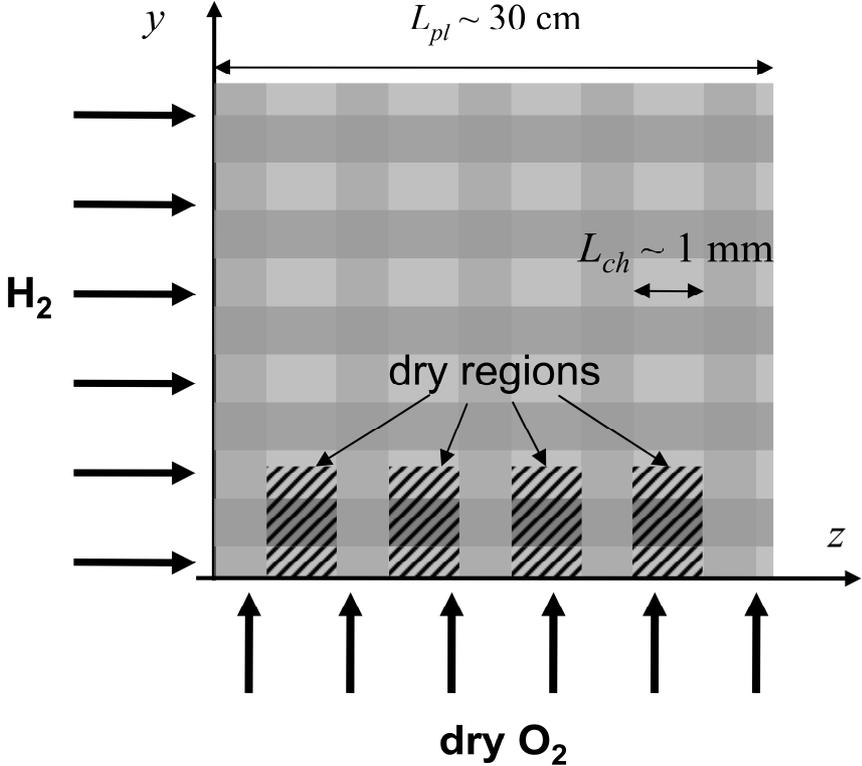

Figure 2

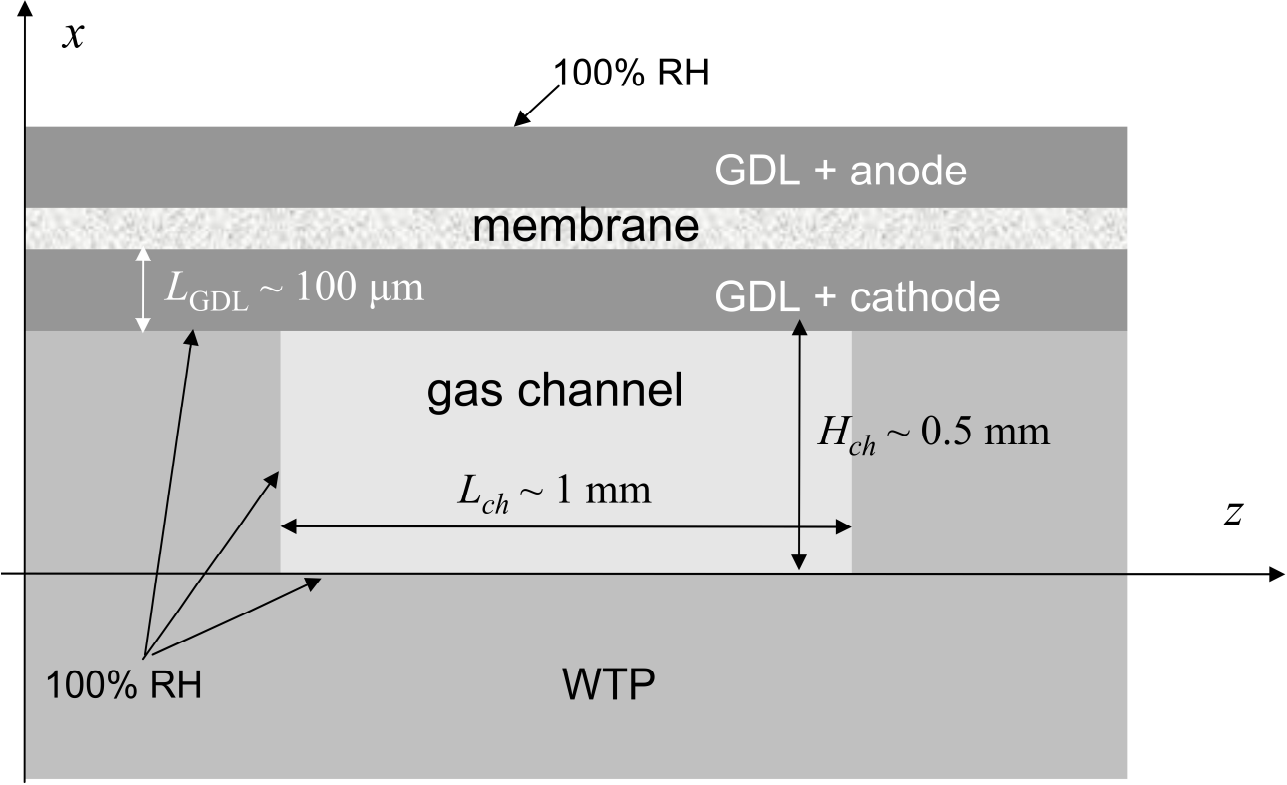

Figure 3

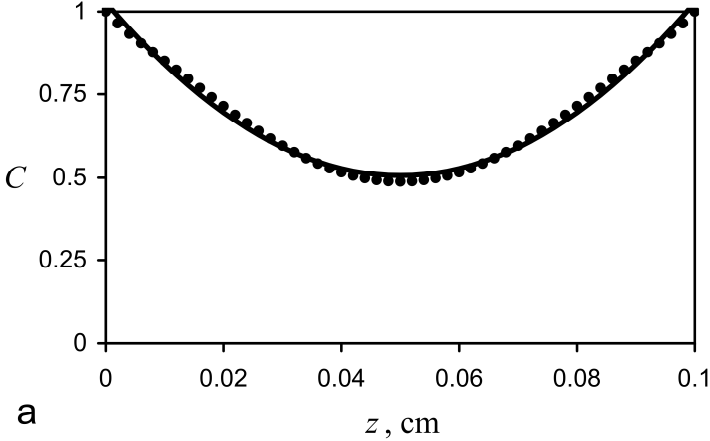

a

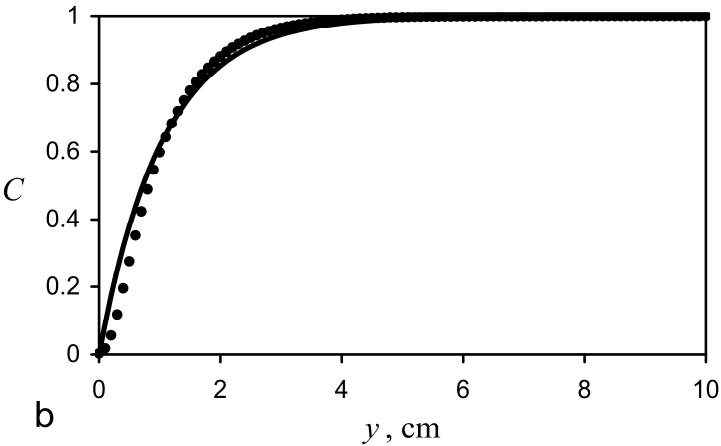

b

Figure 4

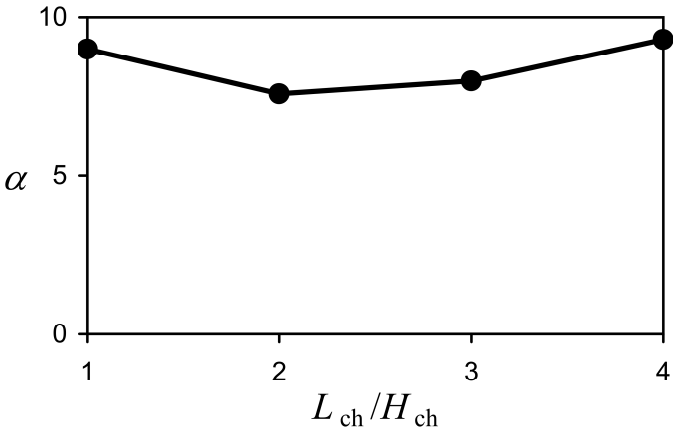

Figure 5

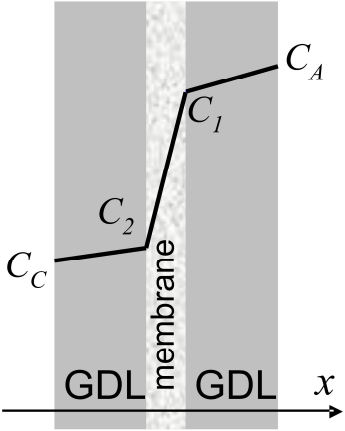

Figure 6

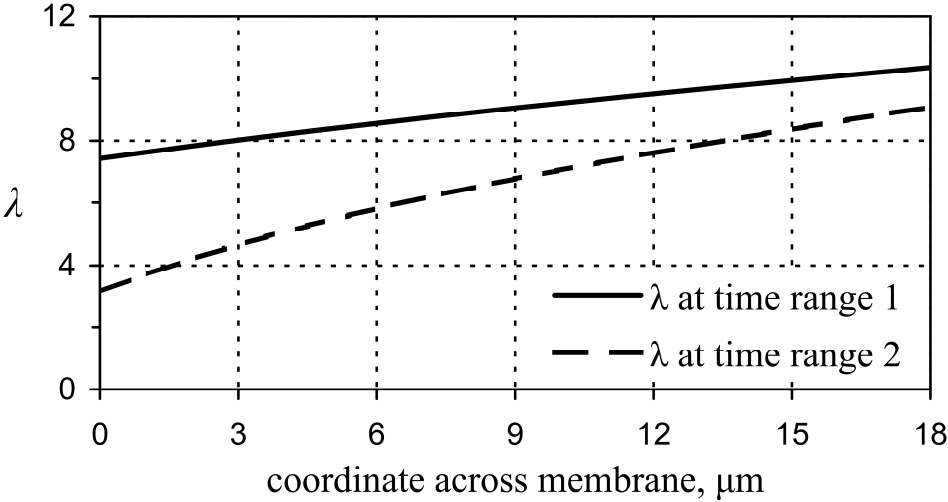

Figure 7

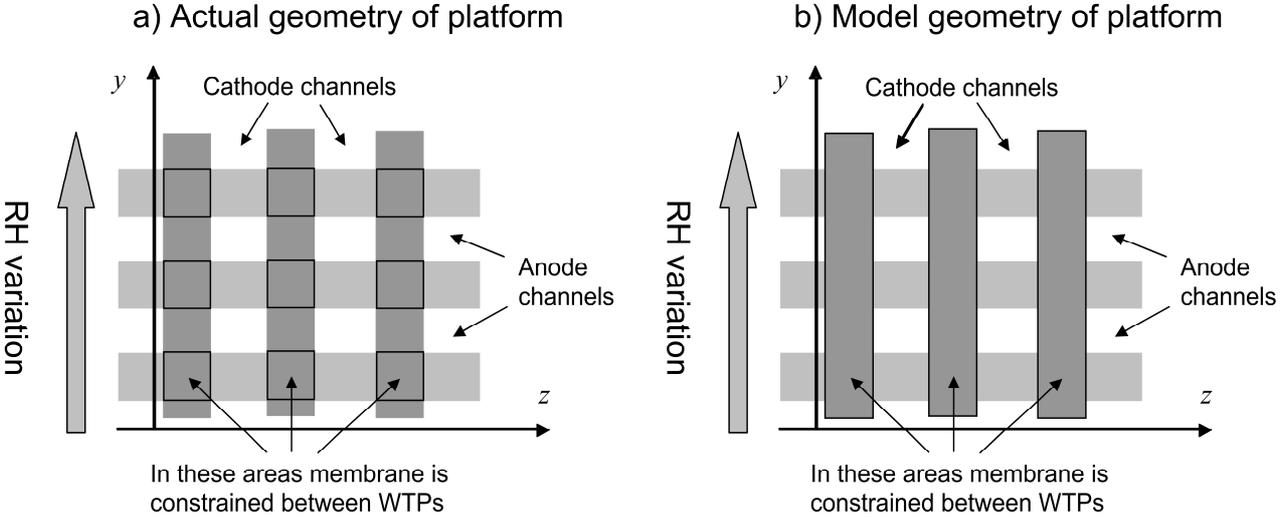

Figure 8

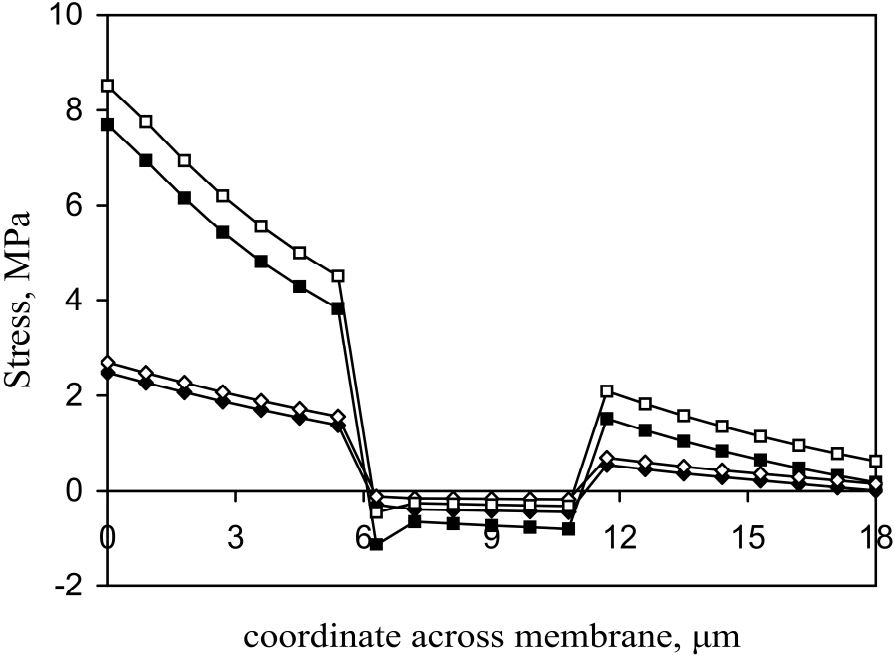

coordinate across membrane, μm

Figure 9

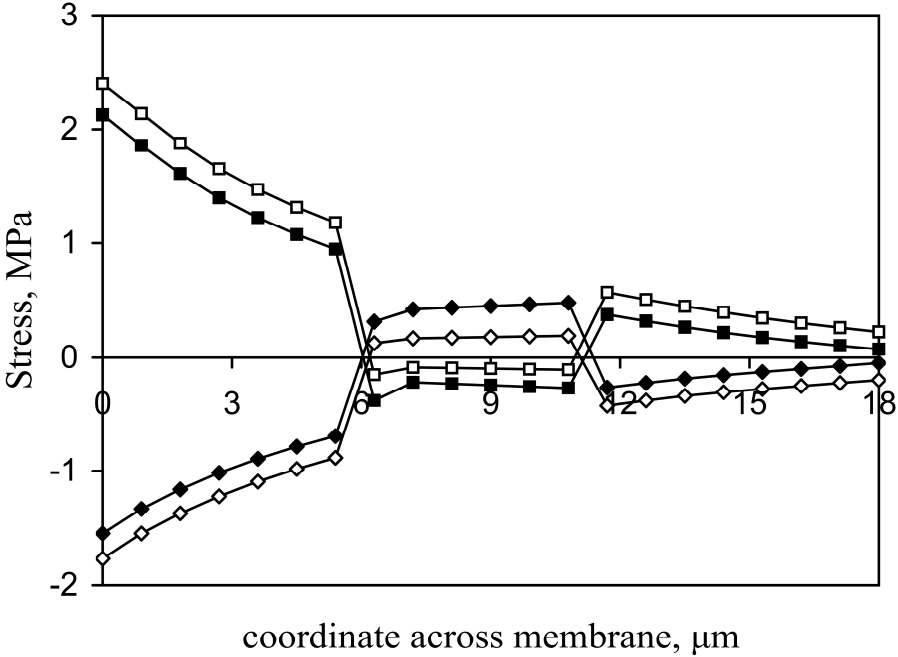

coordinate across membrane, μm

**Figure captions**

**Figure 1.** Qualitative picture of gas flow field and RH distribution in platform.

**Figure 2.** The cross-section of the cathode part of fuel cell.

**Figure 3.** The numerical solution for the vapor concentration at the GDL/gas channel interface (points) and the function $f(y, z)$ (solid line) in the channel cross-section y = 0.5 cm (a), z = $L_{ch}$/2 (b). The following parameter values are used: $L_{ch}$ = 0.1 cm, $H_{ch}$ = 0.05 cm, $V_0$ = 3 m/s, $D$ = 2.4·$10^{-5}$ $m^2$ $s^{-1}$, $C_{in}$ = 0, $C_{sat}$ = 1.

**Figure 4.** Dependence of the fitting parameter α on the ratio $L_{ch}$/$H_{ch}$.

**Figure 5.** Vapor concentration in membrane/electrodes/GDLs assembly cross-section.

**Figure 6.** The through-plane water distribution in the membrane cross-section at y = 5 mm, z = $L_{ch}$/2 for two values of the current density.

**Figure 7.** The actual (a) and the modeled (b) geometries of the membrane constraints. The membrane is constrained between the WTPs through the GDLs in the areas indicated by gray rectangles.

**Figure 8.** The in-plane elastic stress in the membrane ($\sigma_{zz}$, solid symbols, and $\sigma_{yy}$, empty symbols) as a function of through-plane coordinate, $x$, in the membrane cross-section near the air inlet (y = 0.5 cm) at the middle of the channel (x = $L_{ch}$/2) for two current densities, $j$ = 100 mA $cm^{-2}$ (diamonds) and $j$ = 1000 mA $cm^{-2}$ (squares). Model parameters: temperature 75°C, *α=0.036* for ionomer and *α=0* for reinforcement, *v=0.4*, the dependence of *E* on *λ* is taken from ref[31].

**Figure 9.** The in-plane elastic-plastic stress in the membrane ($\sigma_{zz}$, solid symbols, and $\sigma_{yy}$, empty symbols) near the air inlet (y = 0.5 cm) at the middle of the channel (x = $L_{ch}$/2) for two current densities, $j$ = 100 mA $cm^{-2}$ (diamonds) and $j$ = 1000 mA $cm^{-2}$ (squares). Model parameters: temperature 75°C, *α=0.036* for ionomer and *α=0* for reinforcement, *v=0.4*, the dependence of *E* on *λ* is taken from ref[31].